\pdfoutput=1
\documentclass[%
aps,
prl,
superscriptaddress,
twocolumn,
showpacs,
amsmath,
amssymb,
amsfonts,
]{revtex4-1}

\usepackage{graphicx}
\usepackage{color}
\definecolor{lightgray}{gray}{0.7}
\definecolor{cream}{RGB}{222,217,201}
\definecolor{green}{rgb}{0.3,0.6,0.3}
\definecolor{blue2}{rgb}{0.5,0.5,1.}

\usepackage{lipsum}

\begin{document}

\title{
Rheology of dense granular flows for elongated particles
}

\newcommand{\SZFI}{%
Institute for Solid State Physics and Optics,
Wigner Research Center for Physics,
Hungarian Academy of Sciences,
P.O. Box 49, H-1525 Budapest, Hungary}

\newcommand{\ESPCI}{%
Physique et M\'ecanique des Milieux H\'et\'erog\`enes,
PMMH UMR 7636 ESPCI - CNRS - Univ. Paris-Diderot -
Univ. P.M. Curie, 10 rue Vauquelin, 75005 Paris, France}

\author{D\'aniel B. Nagy}
\affiliation{\SZFI}

\author{Philippe Claudin}
\affiliation{\ESPCI}

\author{Tam\'as B\"orzs\"onyi}
\affiliation{\SZFI}

\author{Ell\'ak Somfai}
\email{somfai.ellak@wigner.mta.hu}
\affiliation{\SZFI}

\date{\today}

\begin{abstract}
We study the rheology of dense granular flows for frictionless
spherocylinders by means of 3D numerical simulations.
As in the case of spherical particles, the effective friction $\mu$ is an
increasing function of the inertial number $I$, and we systematically
investigate the dependence of $\mu$ on the particle aspect ratio $Q$, as well
as that of the normal stress differences, the volume fraction and the
coordination number. We show in particular that the quasi-static friction
coefficient is non-monotonic with $Q$: from the spherical case $Q=1$, it first
sharply increases, reaches a maximum around $Q \simeq 1.05$, and then gently
decreases, reaching back its initial value for $Q \simeq 2$. We provide a
microscopic interpretation for this unexpected behavior through the analysis
of the distribution of dissipative contacts around the particles: as compared
to spheres, slightly elongated grains enhance contacts in their central
cylindrical band, whereas at larger aspect ratios particles tend to align and
dissipate by preferential contacts at their hemispherical caps.
\end{abstract}


\maketitle

Rheology of dense granular flows is an active domain of research, motivated by
fundamental questions on this `complex fluid' as well as by practical needs in
soil mechanics and geotechnical engineering. 
Since the 1950s a number of models have been suggested, including Bagnold's
scaling \cite{bagnold-procrsoca-1954,silbert-pre-2001}, the theory of the
rapid flow regime \cite{jenkins-jfluidmech-1983} as well as other regimes
\cite{campbell-jfluidmech-2002}.
A major step in the description of the dense regime
has been achieved 10-15 years
ago with the development of the framework of the now so-called $\mu(I)$
rheology \cite{gdrmidi-eurphysje-2004, forterre-annurevfluidmech-2008}, which
successfully describes these flows in the absence of strong spatial
gradients or temporal changes. This approach has shown that the
constitutive
equations, which augment the conservation laws for a complete rheological
description of these flows, can be formalized in terms of the `inertial
number' $I=\dot\gamma d / \sqrt{p/\rho}$, where $\dot\gamma$ is the shear
rate, $p$ is the pressure, $d$ is the average grain diameter, and $\rho$ is
the density of the particles' material. This dimensionless number can be
interpreted as the ratio of the characteristic time scale $d/\sqrt{p/\rho}$ of
microscopic rearrangements, and the macroscopic time scale $1/\dot\gamma$ of
the deformation. In the case of rigid grains, for which the pressure is the
only stress scale, the dimensional analysis tells us that the effective
friction $\mu$ of the flow, defined by the ratio of the shear stress to the
pressure, as well as the volume fraction $\phi$ of the granular packing, are
functions of $I$. The shape of these functions has been determined both by
simulations \cite{cruz-pre-2005, jop-nature-2006, hatano-pre-2007} and
experiments \cite{jop-jfluidmech-2005, fall-jrheol-2015}.

The $\mu(I)$ formalism has been successfully applied in a number of flow
geometries \cite{gdrmidi-eurphysje-2004}, recovering the Bagnold scaling in
flows down an inclined plane \cite{silbert-pre-2001}, and describing chute
flow \cite{gray-jfluidmech-2014, barker-jfluidmech-2015,
baker-jfluidmech-2016, borzsonyi-prl-2009}, silo discharge
\cite{staron-eurphysje-2014}, granular column collapse
\cite{lagree-jfluidmech-2011, ionescu-jnonnewtonfluidmech-2015}, and dynamic
compressibility effects in spontaneous oscillatory motion
\cite{staron-pre-2012, trulsson-epl-2013}. This rheology has been extended in
a number of ways taking into account various effects, like cohesion
\cite{rognon-epl-2006}, finite pressure or soft particles
\cite{favier-submittedprl-2017}, and self-propelling particles
\cite{peshkov-epl-2016}. Another example for extension is the description
of granular suspensions \cite{boyer-prl-2011, trulsson-prl-2012,
degiuli-pre-2015}.  In this case, a new time scale, that of the viscous
dissipation, is introduced, which is captured by a new dimensionless group,
the `viscous number' $J$. This formalism describes Brownian suspensions as
well \cite{trulsson-epl-2015}, and has been incorporated to diphasic models
for sediment transport \cite{ouriemi-jfluidmech-2009,
aussillous-jfluidmech-2013, chiodi-jfluidmech-2014}.

\begin{figure}[t!]
  \includegraphics[width=70mm]{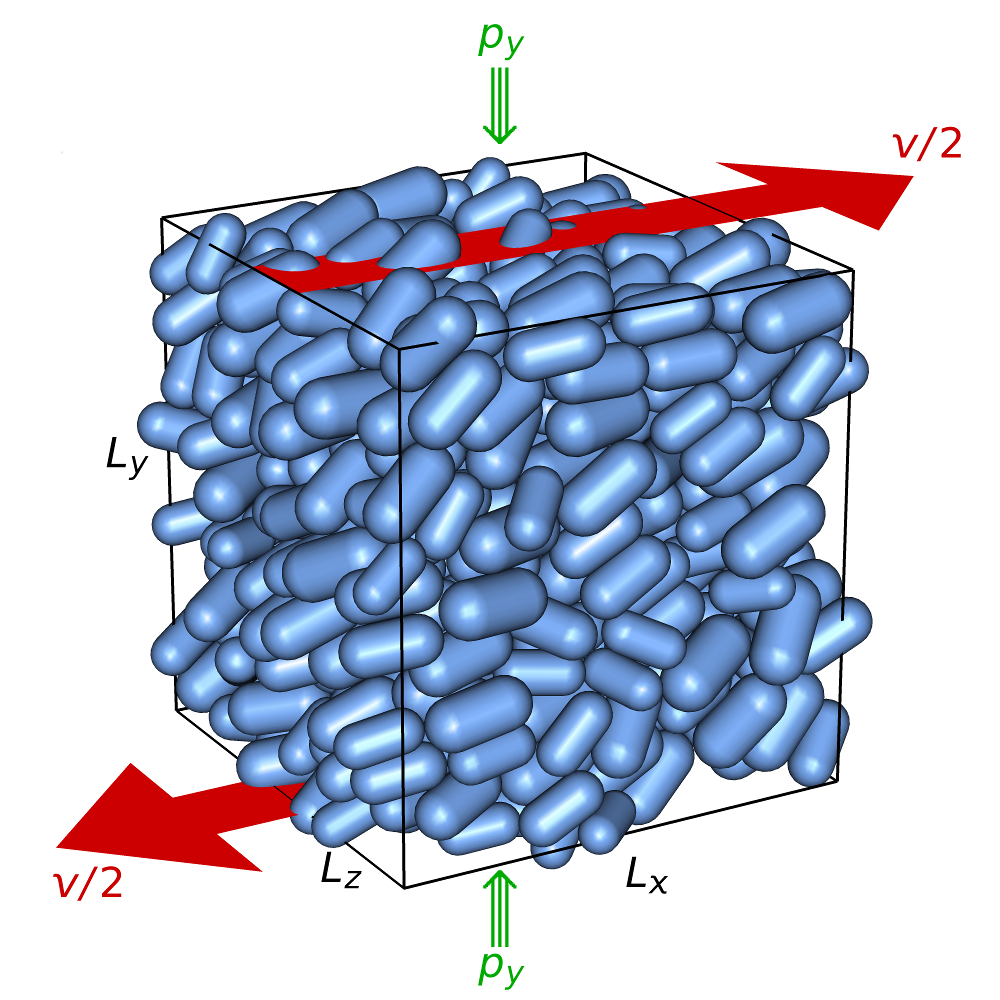}  
  \caption{(Color online)
      The simulation box, here containing $500$ spherocylindrical particles
      of aspect ratio $Q=2$ in the stationary state, is sheared in the $x$
      direction.  A feedback loop adjusts $L_y$ to ensure a controlled applied
      stress $-p_y$ in the $y$ direction. The shear
      rate is $\dot\gamma=v/L_y$.
  }
  \label{fig:box}
\end{figure}

\begin{figure*}[t!]
  \includegraphics[width=178mm]{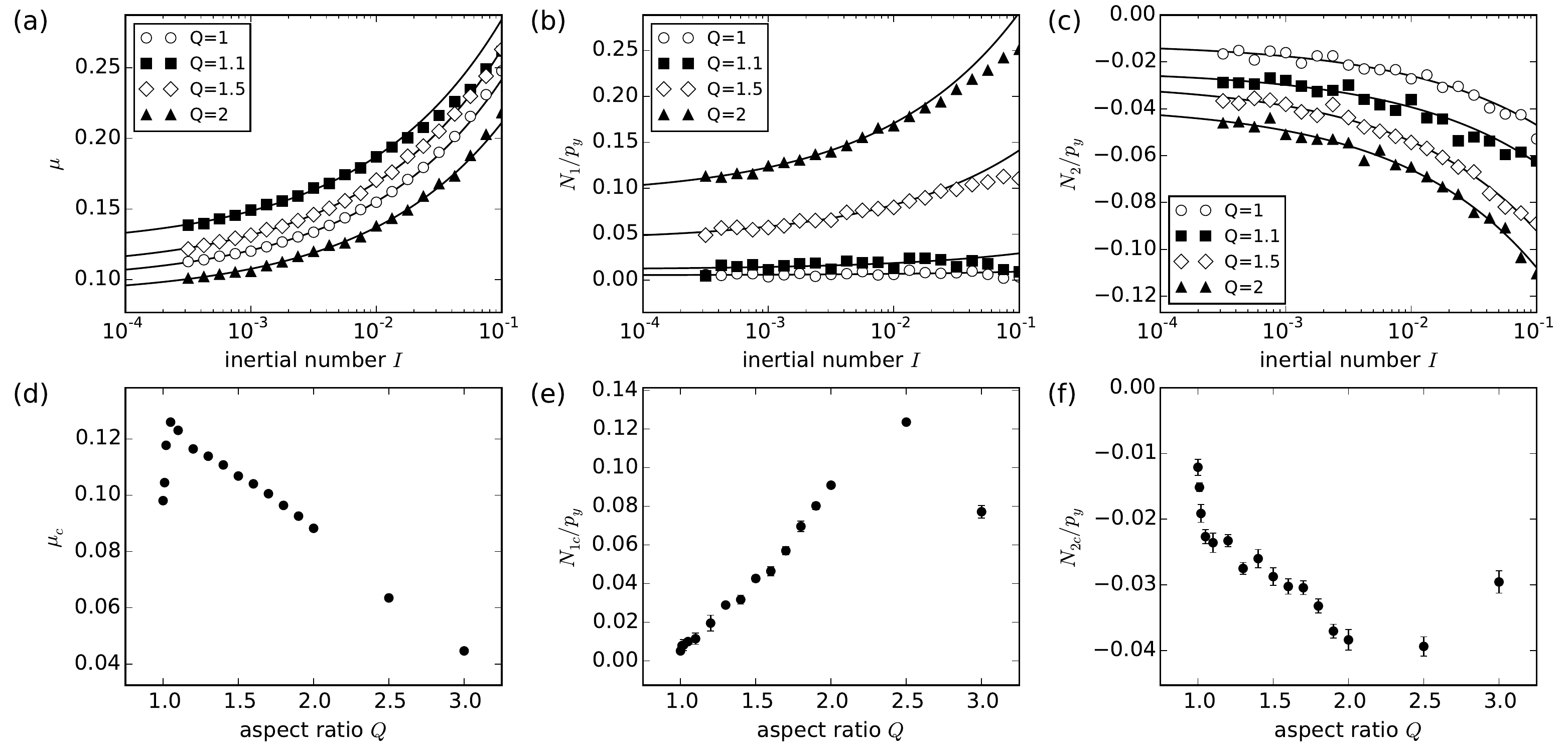}
  \caption{
    Top row: (a) effective friction, (b) first and (c) second normal stress
    differences, as functions of $I$.  The normal stress differences are
    normalized by the absolute value of the imposed stress, $p_y$.  The
    solid curves are fits of the
    form Eq.\ \ref{eq:fit} with $\alpha=0.4$ in the range $10^{-3.5}\le I \le
    10^{-2}$.
    Bottom row: aspect ratio dependence of the quasi-static ($I\to 0$) values
    of the same quantities: (d) effective friction, (e) first and (f) second
    normal stress differences. The first few points correspond to $Q=1, 1.01,
    1.02, 1.05$ and $1.1$.
  }
  \label{fig:rheology}
\end{figure*}

Another trend in granular physics is considering shape anisotropy for the
particles, see the recent review \cite{borzsonyi-softmatter-2013} and
references therein. One of the fundamental results is the observation that
elongated particles get oriented in shear flow
\cite{reddy-pre-2009,campbell-physfluids-2011,borzsonyi-prl-2012,
boton-pre-2013}. The
average orientation angle $\theta_\text{av}$, which is the angle between the
average orientation of the particles and the streamlines, is nonzero; it
decreases with the length-to-width aspect ratio $Q$ of the particles, but only
weakly depends on the shear rate \cite{borzsonyi-pre-2012}. There is an
interplay between this orientational ordering and the packing fraction or the
contact force network \cite{azema-pre-2010, azema-pre-2012,
wegner-softmatter-2014}. The quasi-static behavior of 2D systems has been
investigated with rounded-cap rectangular particles in a biaxial set-up
\cite{azema-pre-2010}. The fast regime (typically $I>0.1$) has been explored
in various 2D configurations, including a volume fraction controlled shear
cell, with dumbbells \cite{reddy-jfluidmech-2010, mandal-physfluids-2016}. In
this paper, we study the rheological properties of assemblies of 3D
frictionless spherocylinders in a pressure-controlled shear cell. We explore
the range of shear rate for which the $\mu(I)$ formalism is expected to apply,
i.e. from the quasi-static limit ($I \to 0$) to the beginning of the kinetic
regime $I \simeq 0.1$. We unexpectedly find a non-monotonic behavior of the
quasi-static friction coefficient with the particle aspect ratio, a key result
missed by previous studies \cite{azema-pre-2010,mandal-physfluids-2016}. We
also report the emergence of normal stress differences, whose marginal
presence was already noticed for 3D flows of spheres \cite{depken-epl-2007},
but which clearly develop for elongated particles in a way qualitatively
similar to those in suspensions of fibers \cite{snook-jfluidmech-2014,
bounoua-jrheol-2016}.
A very recent experimental study of the rheology of non-colloidal
suspensions of rigid fibers has investigated the effect of the particle
aspect ratio in the range $3$--$15$, indicating an aspect ratio
independent friction coefficient, but a decreasing jamming packing
fraction with increasing $Q$ \cite{tapia-jfluidmech-2017}.
A non-monotonic dependence of the packing fraction on $Q$ has been
observed previously also in non-sheared systems \cite{williams-pre-2003,
donev-science-2004, rodney-prl-2005}.

\emph{Numerical setup.}---
To model homogenous shear flow, we used a 3D plane-Couette geometry, with
periodic boundary conditions in the $x$ (flow) and $z$ (neutral) directions,
and Lees-Edwards boundary conditions in the $y$ (velocity gradient) direction
(Fig.~\ref{fig:box}); this way undesirable effects of walls could be
eliminated \cite{artoni-prl-2015}.  Instead of cylinders
\cite{guo-jfluidmech-2012} we chose the spherocylindical shape, because of the
availability of efficient numerical algorithms \cite{vega-computchem-1994} and
continuous transition to the reference spherical shape. The spherocylinders
were parametrized by their length-to-diameter aspect ratio $Q=\ell/2R$. The
repulsive force ${\mathbf F}_{ij}$ between particles $i$ and $j$ was
proportional to their normal overlap, and we employed a viscous
velocity-difference based term for dissipation:
\(
{\mathbf F}_{ij} = (-k\, \delta_{ij}
      + b\, {\mathbf v}_{c,ij}\cdot\hat c_{ij})\, \hat c_{ij} \,,
\)
where $\delta_{ij}$ and $\hat c_{ij}$ are the magnitude and unit direction
vector of the normal overlap between the particles, and for the velocity
difference ${\mathbf v}_{c,ij}$ at contact the rotation of the particles were
taken into account as well. There is no tangential component of the force,
that would result from a Coulombic contact friction. The stiffness $k$ of the
contacts, the particle diameter $2R$ and density $\rho$ were set to unity,
implicitly defining the length, time and mass units of the simulation.
Importantly, some polydispersity is introduced to reduce the effects of
crystallization at large $Q$, and while we kept the aspect ratio constant, we
have drawn the radii of the particles from a uniform distribution with a ratio
of standard deviation to mean of 10\%. The prefactor $b$ in the dissipative
term was set by specifying a given restitution coefficient for binary
collision. The equations of motion were integrated by the velocity-Verlet
scheme, representing particle rotations by quaternions
\cite{omelyan-computphys-1998}.

\begin{figure*}
  \includegraphics[width=130mm]{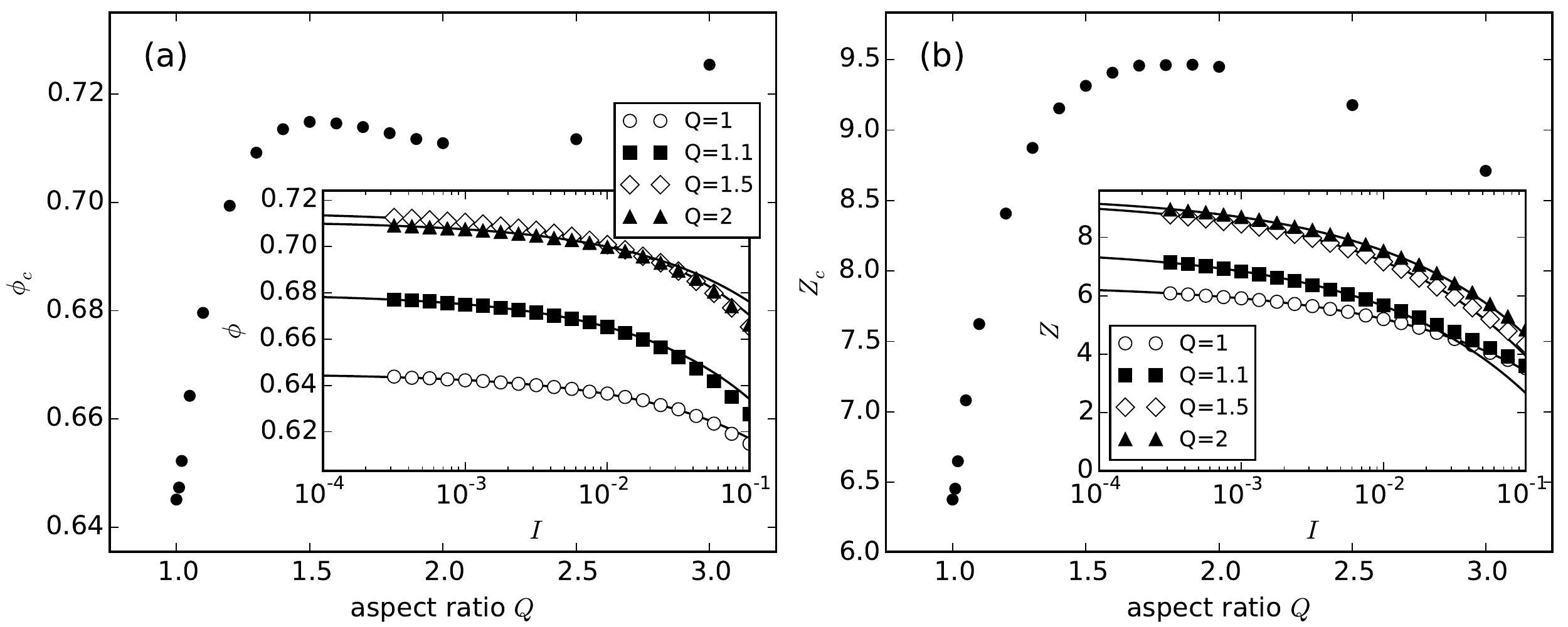} 
  \caption{
    The aspect ratio dependence of the quasi-static values of (a) the volume
    fraction, and (b) the coordination number.  Insets: $I$ dependences for a
    few $Q$ values. Solid lines: fits of the form of Eq.~\ref{eq:fit}, with
    respective fixed exponents $0.4$ and $0.5$.
  }
  \label{fig:phiz}
\end{figure*}

We created the initial conditions of random particle orientation with
overdamped dynamics, and afterwards sheared the system at constant shear rate
$\dot\gamma$. During shear we employed stress control, where one side of the
box, $L_y$ was adjusted by a feedback loop such that the corresponding normal
stress $\sigma_{yy}$ fluctuated around a specified value
$-p_y$. We have used
$p_y=10^{-3}$ in these units, corresponding to the
rigid limit where rescaled results become independent of
$p_y$. We kept $L_x$ and $L_z$ fixed in order to
avoid the development of a singular box shape due to normal stress
differences. All measurements were taken in the stationary state, reached
after a deformation of $\gamma=25$ when starting from the initial conditions,
or of $\gamma=10$ from the stationary state of a different shear rate. All
quantities of interest were time-averaged at least over an additional
deformation of $\gamma=5$. In the steady state, the packing fraction as well
as all stress components $\sigma_{ij}$ were homogenous, and the velocity
profile linear (no shear banding). Finally, we checked that our results are
independent of the integration time step, set to $1/100$ of the duration of a
binary collision, and qualitatively insensitive to the value of the
restitution coefficient in the range 0.3--0.7, and here set to $0.5$.

\emph{Results.}---
We measured the inertial number dependence of different quantities. In
Fig.~\ref{fig:rheology}(a) we show the effective friction,
$\mu=\sigma_{xy}/p_y$
vs $I$ for spherocylinders of a few selected aspect ratios, and for spheres
for reference. (Note that in the definition of $I$ we used the value
$p_y$ controlling the stress, instead of the pressure $p$.) Similarly to
\cite{degiuli-pre-2015}, we fitted the empirical form
\begin{equation}
  \mu(I) \approx \mu_c + \mu_1 I^\alpha\,,
  \label{eq:fit}
\end{equation}
allowing us to extrapolate to the quasi-static friction $\mu_c$ in the $I \to
0$ limit. Best fitting is of course obtained when all three parameters are
adjusted, but we then fixed the exponent to its average, $\alpha=0.4$,
yielding less noisy data with a two parameter fit ($\mu_c$ and $\mu_1$). This
value of $\alpha$ is in agreement with that deduced from 3D simulations of
frictionless hard spheres \cite{peyneau-pre-2008}, and similar to the exponent
$0.5$ observed for frictionless circles \cite{bouzid-prl-2013}.
Fig.~\ref{fig:rheology}(d) displays the aspect ratio dependence of $\mu_c$,
showing a surprising non-monotonic function: it rises steeply for aspect
ratios slightly larger than one, takes the highest value around $Q=1.05$,
followed by a slow decrease -- we shall give a microscopic interpretation of
this behavior later. Similarly, we show the normalized first and second normal
stress differences $N_1/p_y = (\sigma_{xx} - \sigma_{yy}) / p_y$
and $N_2/p_y =
(\sigma_{yy} - \sigma_{zz}) / p_y$ in Fig.~\ref{fig:rheology}(b-c). Their
quasi-static values, extrapolated with a fit like Eq.~(\ref{eq:fit}) (also with
fixed $\alpha=0.4$), are shown in Fig.~\ref{fig:rheology}(e-f). As expected,
the quasistatic first normal stress difference vanishes for spheres, and
increases close to linearly for aspect ratios $Q \lessapprox 2.5$. The second
normal stress difference is negative although 2--3 times smaller in amplitude
than $N_1$ (even for spheres $N_{2c}/p_y \simeq 1$\%), and has a very strong
aspect ratio dependence for nearly spherical particles.

In complement to the stress behaviors, we show the volume fraction in
Fig.~\ref{fig:phiz}(a). As for spheres, $\phi$ decreases with $I$ as larger
shear rates generate a more dilute system. Also, consistently with what has
been observed for spherocylinders \cite{williams-pre-2003} and ellipsoids
\cite{donev-science-2004} in non-sheared systems, the volume fraction of
elongated particles first quickly increases with $Q$, followed by a slow
decrease -- the behavior beyond $Q\gtrapprox 2.5$ will be discussed later.
Finally, we computed the coordination number $Z$. A static packing of
frictionless hard spheres takes the isostatic value $6$. Once the particles
become elongated, due to the two extra rotation degrees of freedom per
particle, this value jumps to $10$. In Fig.~\ref{fig:phiz}(b) we plot the
measured values. For spheres $Z_c$ is slightly larger than $6$ due to the
finite pressure, or equivalently the softness of the particles. For increasing
aspect ratio $Z_c$ increases initially sharply but continuously, reaching a
flat maximum around $Q\approx 1.8$. The inset shows the inertial number
dependence, demonstrating an expected decrease of contacts for more violent
flows.

\begin{figure}[t!]
  \includegraphics[width=80mm]{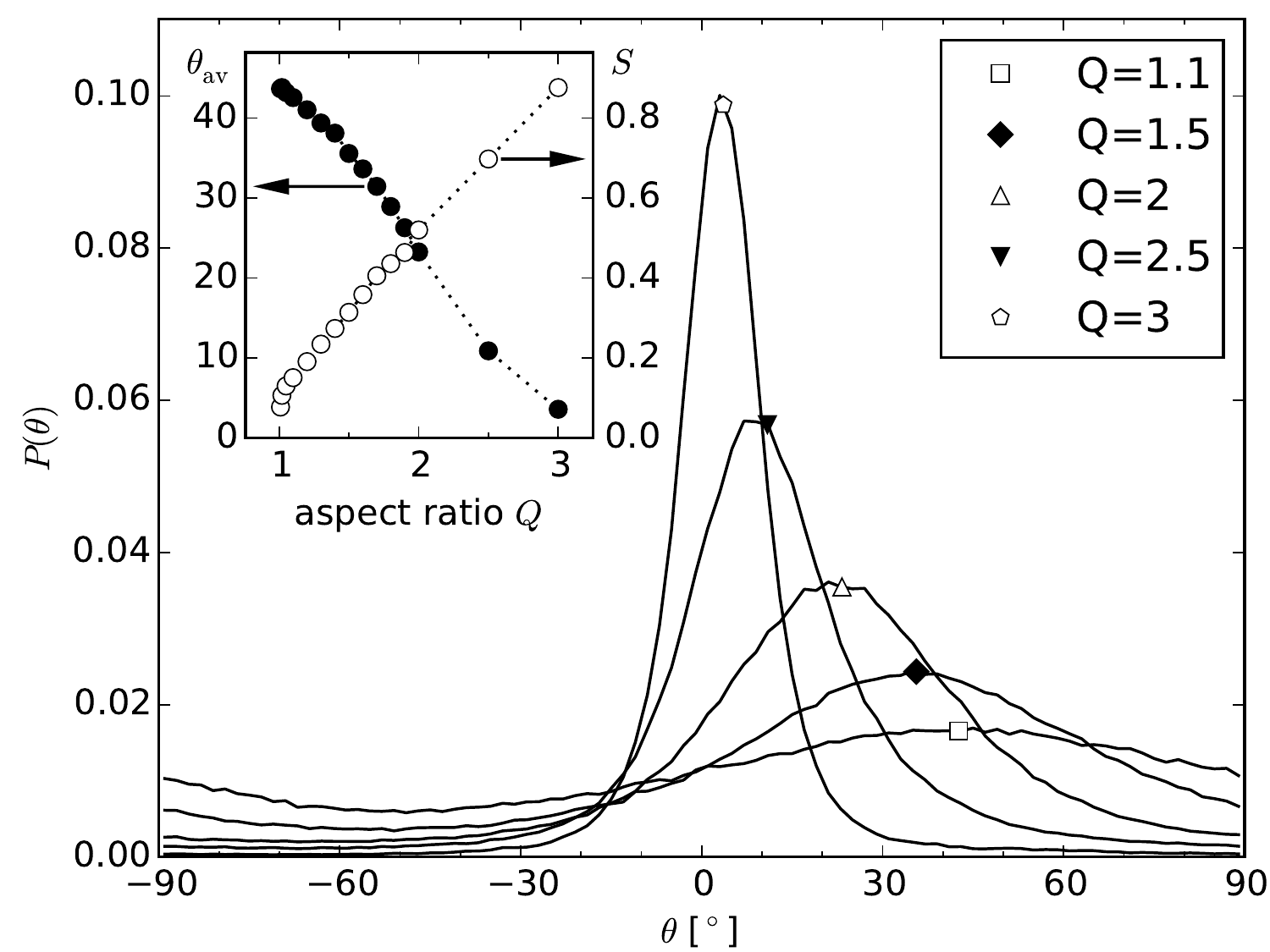}
  \caption{
    Orientation distributions for five different aspect ratios (see legend).
    The symbols locate the average orientation $\theta_{\rm av}$.
    Inset: average orientation $\theta_{\rm av}$ (black bullets, left axis)
    and nematic order parameter $S$ (withe circles, right axis) as functions
    of $Q$.
    These data are for $I=3.16\times 10^{-4}$.
  }
  \label{fig:orient}
\end{figure}

Next we look at the orientational order induced by the shear deformation.
Figure~\ref{fig:orient} shows, for a few selected aspect ratios, the
distribution of the angle $\theta$ between the streamlines ($x$ axis) and the
projection of the particle axis onto the $x$-$y$ shear plane. For more
elongated particles the distribution is narrower and its mode and mean are
closer to zero. The average angle $\theta_\text{av}$ is slightly off the mode
due to asymmetry of the distribution. The aspect ratio dependence of
$\theta_\text{av}$ and the nematic order parameter $S$, which is defined as
the largest eigenvalue of the order tensor, are shown in the inset: increasing
$Q$ results in increasing $S$ and reducing $\theta_\text{av}$, in almost
linear fashions. Note that even the smallest aspect ratio considered,
$Q=1.01$, has a small but finite nematic order, with $\theta_\text{av} \simeq
45^\circ$.

\begin{figure}[t!]
  \includegraphics[width=80mm]{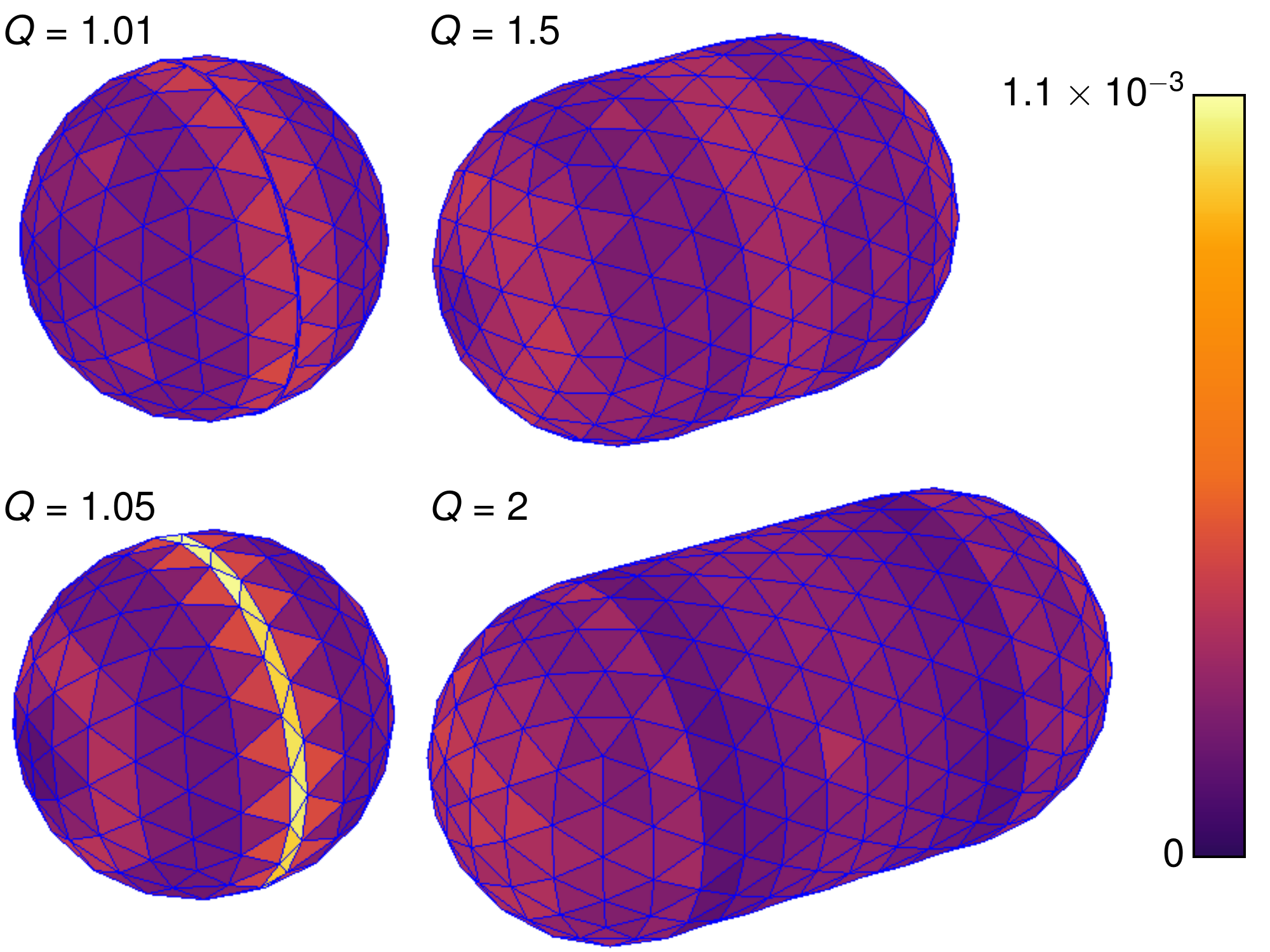}
  \caption{(Color online)
    Average dissipative power per unit area visualized on a discrete mesh for
    four aspect ratios (see legends and color code). Dissipation at each
    contact is accumulated in triangular bins, such that the axes of the
    differently oriented particles are turned into a canonical orientation, in
    which the mesh is defined.
    These data are for $I=3.16\times 10^{-4}$.
  }
  \label{fig:mesh}
\end{figure}

Finally we determined which region of the particles' surface is most
responsible for dissipation (Fig.~\ref{fig:mesh}). The highest dissipation
density was observed in the cylindrical band of the $Q=1.05$ spherocylinder.
More spherical particles showed a more homogenous distribution, but the slight
enhancement of the dissipation near the cylindrical band is visible even for
$Q=1.01$. For more elongated particles, the dissipation becomes dominated at
the hemispherical caps.

\emph{Discussion and perspectives.}---
The framework of the $\mu(I)$ rheology can be extended to elongated particles.
The aspect ratio dependence of the rheological quantities display two
remarkable features: (i) the dissipation, quantified by $\mu_c$, is maximal
around $Q=1.05$, and (ii) the normal stress differences, the volume fraction
as well as the coordination number, behave non-monotonically for $Q\gtrapprox
2.5$.

Issue (i) is closely related to the highest observed density of dissipation in
the cylindrical band of the $Q=1.05$ particles. We suspect that this effect is
particularly strong for spherocylindrical particles, where the surface is not
analytic (one of the curvatures is not continuous). An analysis similar to
what is shown in Fig.~\ref{fig:mesh} revealed that not only the dissipation
density, but also the contact density is increased in that region: a larger
than expected number of particle-pairs locked in a configuration where the
contact is carried by the cylindrical region for at least one of the
particles. We explain issue (ii) by the high nematic and partial spatial order
observed for sufficiently elongated frictionless spherocylinders. While the
introduction of polydispersity destroyed the crystalline order perpendicular
to the streamlines, we still observe short range chains of particles which
follow each other on a streamline. This can also explain why the tips of the
particles become the dominant location for dissipation for the more elongated
particles.

The robustness of these results should be assessed by complementary
simulations, e.g. in 2D, with other shapes like ellipses, possibly including
frictional contacts, as well as by experiments \cite{tapia-jfluidmech-2017,
trulsson-preprint-2017}. A better understanding of the
large $Q$ behavior probably requires the use of larger systems, and here our
data for $Q=3$ may be affected by a too small size. This issue motivates
further studies towards the flow of fibers and entangled materials
\cite{rodney-prl-2005}, whose rich mechanics suggest interesting rheological
properties.

\begin{acknowledgments}
We thank A.~Favier de Coulomb, D.~K\'alm\'an, B.~Andreotti, E.~Cl\'ement,
O.~Pouliquen, A.~Lindner, B.~Szab\'o and M.~Trulsson for useful discussions.
This work was supported by the Hungarian National Research, Development and
Innovation Office NKFIH under grant OTKA K 116036, and the Hungarian-French
bilateral T\'eT/Balaton exchange programme. We acknowledge computational
resources provided by NIIF in Hungary at Debrecen.
\end{acknowledgments}

\bibliography{paper}

\end{document}